\documentclass[twocolumn,showpacs,preprintnumbers,amsmath,amssymb]{revtex4}

\usepackage{graphicx}
\usepackage{dcolumn}
\usepackage{bm}

\begin{document}

\preprint{APS/123-QED}

\title{Scaling Law for Radius of Gyration and Its Dependence on Hydrophobicity}

\author{Liu Hong}
 \email{hong-l04@mails.tsinghua.edu.cn}
\author{Jinzhi Lei}%
\affiliation{%
Zhou Pei-Yuan Center for Applied Mathematics, Tsinghua University,
Beijing, P.R. China, 100084
}%

\date{\today}

\begin{abstract}
Scaling law for geometrical and dynamical quantities of biological
molecules is an interesting topic. According to Flory's theory, a
power law between radius of gyration and the length of homopolymer
chain is found, with exponent 3/5 for good solvent and 1/3 for poor
solvent. For protein in physiological condition, a solvent condition
in between, a power law with exponent $\sim2/5$ is obtained. In this
paper, we present a unified formula to cover all above cases. It
shows that the scaling exponents are generally correlated with
fractal dimension of a chain under certain solvent condition. While
applying our formula to protein, the fractal dimension is found to
depend on its hydrophobicity. By turning a physical process-varying
hydrophobicity of a chain by amino acid mutation, to an equivalent
chemical process-varying polarity of solvent by adding polar or
nonpolar molecules, we successfully deprive this relation, with
reasonable agreement to statistical data. And it will be helpful for
protein structure prediction. Our results indicate that the protein
may share the same basic principle with homopolymer, despite its
specificity as a heteropolymer.

\end{abstract}

\pacs{Valid PACS appear here}
\maketitle

\section{Introduction}
It is well known that a protein can refold to its native structure
from denatured state under physiology condition. However, the
mechanism underlying is still unknown and becomes one of basic
intellectual challenges in molecular biology\cite{1}. In the study
of protein folding, radius of gyration, defined as
$R_g=\sqrt{\frac{1}{N}\sum_{i=1}^{N}(\vec R_i-<\vec R>)^2},\; <\vec
R>=\frac{1}{N}\sum_{i=1}^{N}\vec R_i$, is introduced as an important
quantity. It is not only able to describe the static compactness of
a protein structure, but also the folding process from denatured
state to native state. Experimentally, Takahashi \textit{et. al.}
used small-angle X-ray scattering method to measure time evolution
of $R_g$ during a protein's folding process. In their study,
significant changes in radius of gyration from unfolded to folded
conformations were observed in several proteins by pH jump\cite{2}.

An interesting question is about the relationship between $R_g$ and
other physical quantities. In this paper, we present a scaling law
between radius of gyration and the length of protein chain ($N$) by
exploiting Protein Data Bank: $R_g\propto N^{\nu}$, which has also
been reported by other authors\cite{8,9,10,11,12}. Through
generalizing former Flory's theory\cite{3}, we get a new unified
formula, which can be applied to polymer in poor solvent, polymer in
good solvent and protein under physiological condition etc. It shows
that the scaling exponents are generally correlated with the fractal
dimension of a chain. We also study the influence of hydrophobicity
on compactness of a protein chain. By considering the equivalence
between protein-solvent coupled systems, we turn a physical
process-varying hydrophobicity of a chain by amino acid mutation, to
a chemical process-varying polarity of solvent by adding polar or
nonpolar molecules. This enables us to derive a relation between
hydrophobicity and fractal dimension, with good agreement to
statistical data.

The paper is organized as follows: In Section II, a scaling law of
radius of gyration for proteins under physiological condition is
presented. In section III, we deprive our new unified formula based
on Flory's original theory. In Section IV, the influence of
hydrophobicity on fractal dimension is studied. Section V will be a
brief conclusion. In Appendix, the relation between scaling exponent
and hydrophobicity is studied directly, in the same way as Section
IV.

\section{Scaling exponent for protein under physiological condition}
If neglect minor differences between amino acids, protein can be
treated as a homopolymer. According to well-known Flory's
theory\cite{3,4,5,6}, there exists a universal scaling law between
radius of gyration and the length of polymer chain.
\begin{equation}
R_g \propto N^\nu,
\end{equation}
where exponent $\nu$ depends on solvent condition. Under good
solvent condition, monomers are separated by solvent molecules. Thus
we have $\nu=3/5$. Under poor solvent condition, the chain is highly
compressed by solvent pressure. And $\nu = 1/3$ is as high as
crystals.

\begin{figure}[h]
\includegraphics[width=9.0cm]{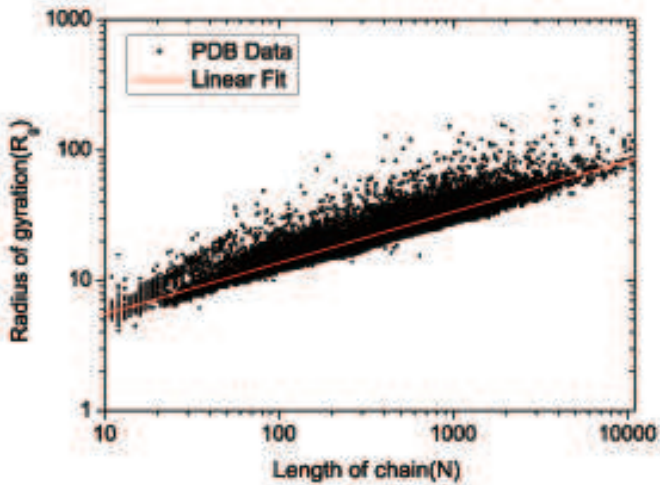}
\caption{A log-log plot of 37162 protein structures in PDB, with
$\nu=0.3916\pm0.0008$ by least-quare linear fit.} \label{fig:1}
\end{figure}

However, proteins under physiological condition have their
specificity. On one hand, they are compact due to hydrophobic
interactions. On the other hand, they are usually not well-packed
and contain many cavities inside\cite{7}. Geometrically, these
cavities are a consequence of regular secondary structures in folded
proteins. Furthermore, they are essential for biological functions,
since they can serve as binding sites when contacting to other
molecules. Therefore, folded proteins should be more compact than
polymers in good solvent, and looser than highly compressed polymers
in poor solvent, i.e., $1/3\leq\nu\leq3/5$. This argument is
confirmed by statistical study of over 37,000 protein structures
from Protein Data Bank (PDB), which yields $\nu \approx 2/5$
(Fig.\ref{fig:1}) and agrees with the research of
Arteca\cite{8,9,10}. It indicates that proteins in native state are
not so compact as crystals, which is a bit different from current
popular view\cite{11}.

The influence of secondary structure is also studied. In Fig. 2, we
show statistical data for all-$\alpha$, all-$\beta$, $\alpha/\beta$
mixed protein structures (the fractions of amino acids in secondary
structures are larger than $50\%$) and unstructured proteins (the
fraction of amino acids in secondary structures is less than $20\%$)
in PDB. Despite their great differences in secondary structure,
their scaling exponent $\nu$ are all approximate to $2/5$. This
result seems a bit contradictive to our common sense at first
glance. Since it is easily to see that for single straight
$\alpha$-helix, $\nu=1$; for perfect planar $\beta$-sheet,
$\nu=1/2$. They are both largely apart from $\nu=2/5$. However, as
secondary structure is a local characteristic, while scaling
exponent $\nu$ mainly depends on over-all topological properties.
When the protein is large enough to contain sufficient secondary
structures, their influence will be quite limited. These results
imply that there may exist a unified mechanism for the scaling law
between radius of gyration and the length of protein chain.

\begin{figure}[h]
\begin{minipage}[t]{0.25\textwidth}
\centering(a)
\includegraphics[width=1.8in]{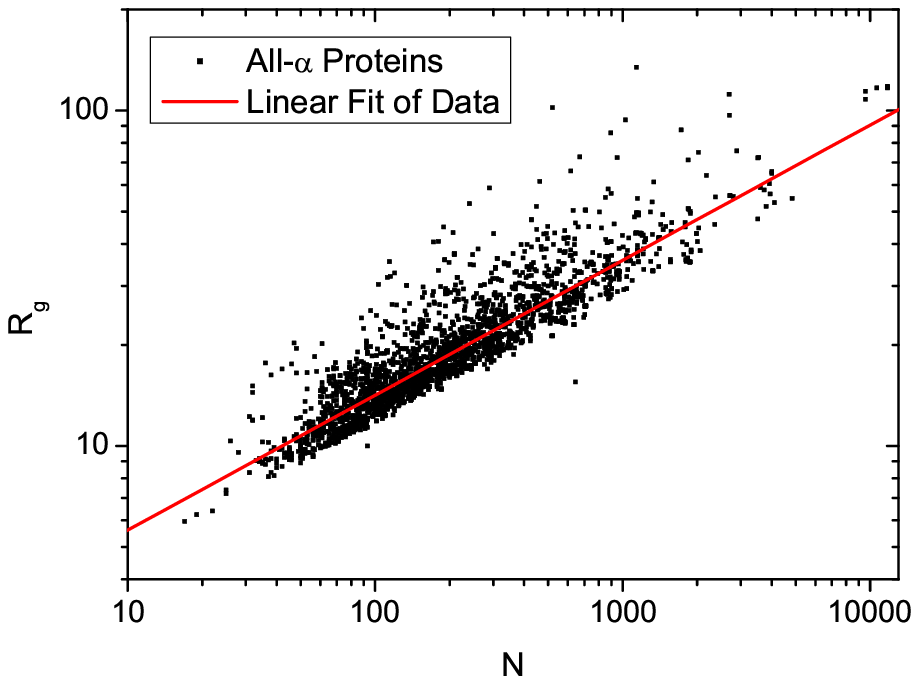}
\end{minipage}%
\begin{minipage}[t]{0.25\textwidth}
\centering(b)
\includegraphics[width=1.8in]{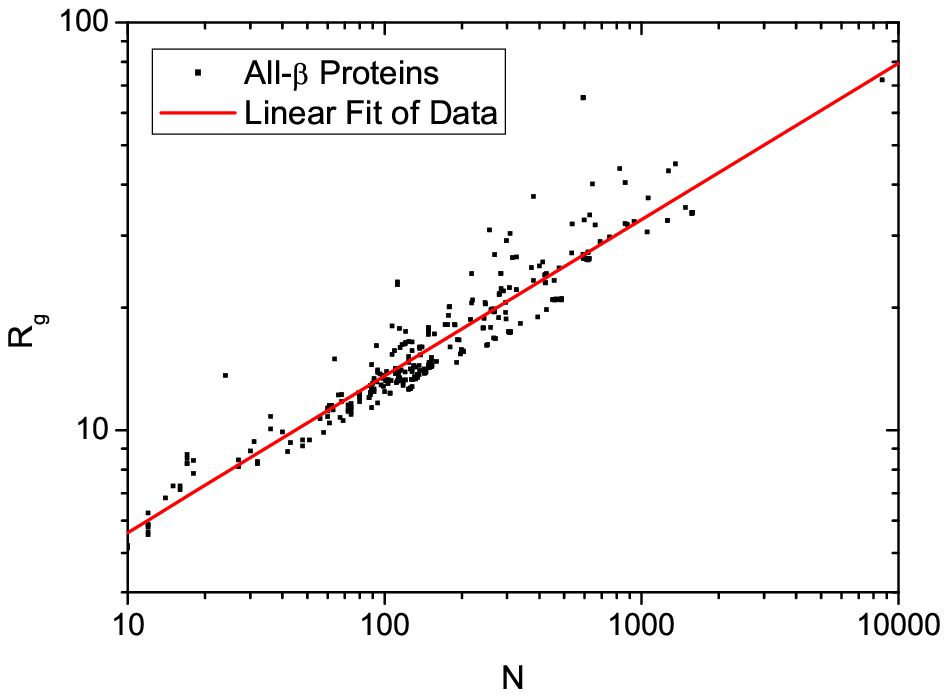}
\end{minipage}
\begin{minipage}[t]{0.25\textwidth}
\centering(c)
\includegraphics[width=1.87in,height=1.32in]{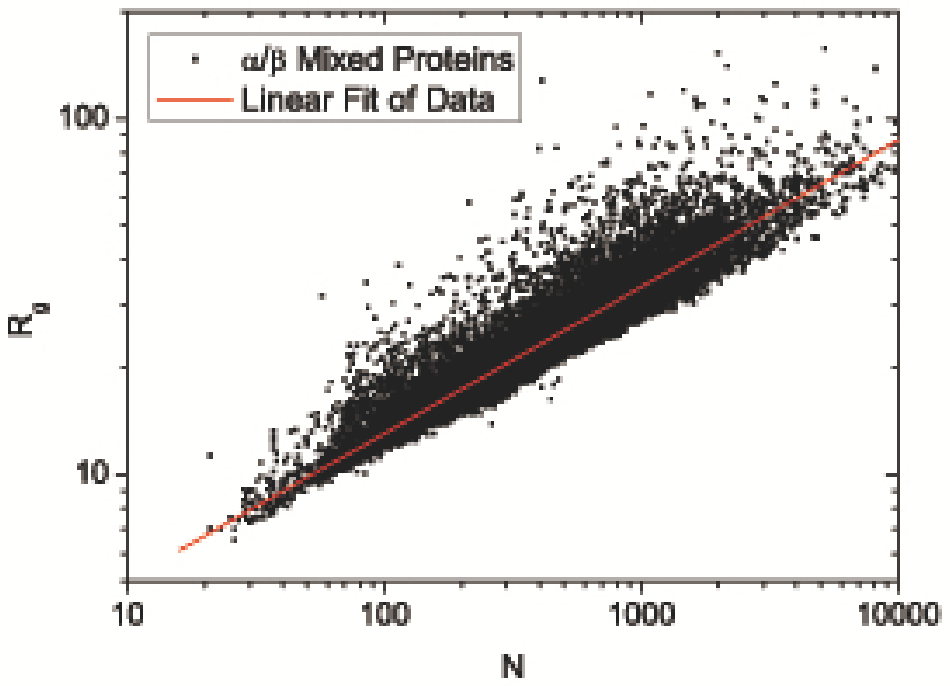}
\end{minipage}%
\begin{minipage}[t]{0.25\textwidth}
\centering(d)
\includegraphics[width=1.8in]{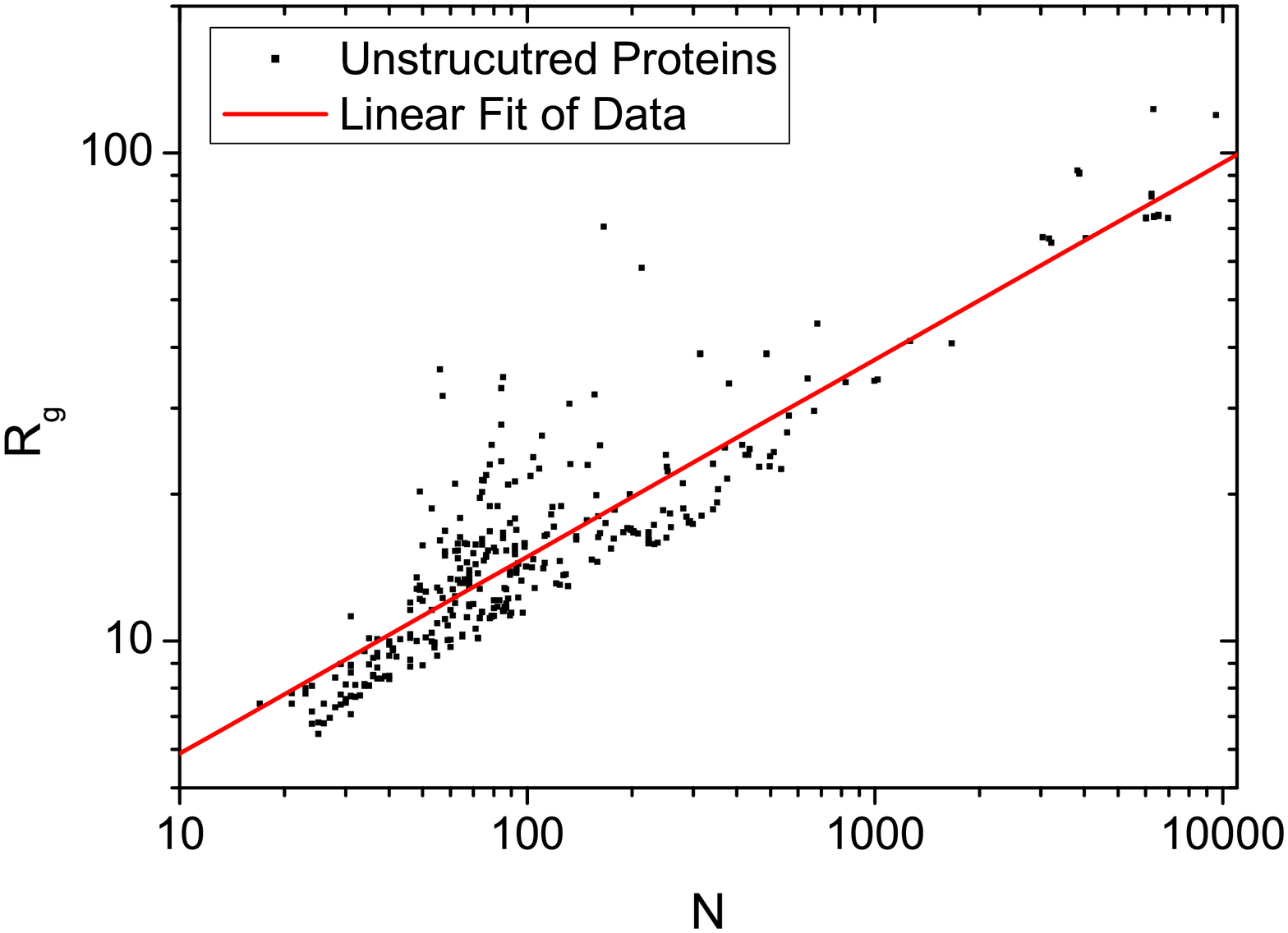}
\end{minipage}
\caption{Log-log plot of proteins with different secondary
structure. (a) 3080 all-$\alpha$ proteins with $N_{\alpha}/N\geq0.5$
($N_\alpha$ is the number of amino acids in $\alpha$-helix. And
single $\alpha$-helix is excluded.). $\nu=0.4026\pm0.0036$ by
least-quare linear fit. (b) 334 all-$\beta$ proteins with
$N_{\beta}/N\geq0.5$ ($N_\beta$ is the number of amino acids in
$\beta$-sheet). $\nu=0.3838\pm0.00746$. (c) 25804 $\alpha/\beta$
mixed proteins with $(N_{\alpha}+N_{\beta})/N\geq0.5$.
$\nu=0.4166\pm0.0010$. (d) 839 unstructured proteins with
$(N_{\alpha}+N_{\beta})/N\leq0.2$. $\nu=0.4038\pm0.0097$.}
\end{figure}

\section{Generalized Flory's Theory}
To obtain a unified formula for scaling law valid under different
solvent conditions, we try to generalize Flory's original
theory\cite{3,4,5,6}. We assume that the chain is made up of N
monomers, which are indistinguishable from each other. Then its
overall size is mainly determined by two following effects: excluded
volume effect that tends to swell the chain, and elastic interaction
that tends to shrink the chain.

Firstly, the excluded volume effect is a consequence of repulsive
interactions between monomers, with energy (two-body repulsive
interaction) given by\cite{3,4,5,6}
\begin{eqnarray}
E_{\mathrm{rep}} = k_B T v \frac{N^2}{R_g^3},
\end{eqnarray}
where $v$ is single monomer's volume.

Then, we calculate the elastic energy. Generally speaking, this term
is originated from contact interactions between monomers, which
include hydrophobic interaction between monomers and solvent
molecules, covalent bonds, hydrogen bonds and Van der Waal's
interaction between neighboring monomers, etc. Since we are unable
to give an explicit formula, we adopt harmonic approximation to find
the dominant part.

Let $d_{ij}$ be the real distance between monomers $i$ and $j$. Then
monomer $i$ is considered to be in contact with monomer $j$, if
$d_{ij}\leq\delta$, where $\delta>0$ is some given constant. Let
$d_0$ be average distance between any two contact monomers $i$ and
$j$. $d_0$ is independent to index $i$ and $j$, and corresponds to
the minimum of potential energy. Under harmonic approximation, the
elastic energy of a chain with $N$ monomers is given by
\begin{eqnarray}
&&E_{\mathrm{ela}} = \frac{1}{2}\sum_{i,j = 1}^{N}\frac{1}{2}\kappa (d_{ij} - d_0)^2\chi(d_{ij}),\nonumber\\
&&\chi(d_{ij})= \left\{ \begin{array}{ll}
1, & \textrm{if $i\neq j$ and $d_{ij}\leq\delta$}\\
0, & \textrm{else}\\
\end{array} \right.\nonumber
\end{eqnarray}
where the first factor $1/2$ dues to double counting of monomers.
And $\kappa$ is Hooke coefficient. Define root-mean-square contact
distance ($d$) as
\begin{eqnarray}
d^2 =\frac{1}{\overline{n}}
\sum_{j=1}^{N}d_{ij}^2\chi(d_{ij}),\nonumber
\end{eqnarray}
where $\overline{n}=\sum_{j=1}^{N}\chi(d_{ij})$ is local contact
number. $\overline{n}$ and $d$ are supposed to be independent to
index $i$, for all monomers are equal in our treatment. On the other
hand, we have
\begin{eqnarray}
\sum_{j=1}^{N}d_{ij}\chi(d_{ij})=\overline{n}d_0.\nonumber
\end{eqnarray}
So now we can rewrite $E_{ela}$ as
\begin{eqnarray}
E_{\mathrm{ela}} = \frac{1}{4}\sum_{i=1}^N \kappa \overline{n}
(d^2-d_0^2)=\frac{1}{4}\overline{n}N\kappa
d^2-\frac{1}{4}\overline{n}N\kappa d_0^2,\nonumber
\end{eqnarray}
As the second term is independent of $R_g$, it will be omitted in
later discussions. Thus, we get
\begin{eqnarray}
E_{\mathrm{ela}} = \frac{1}{4} \kappa \bar{n} N d^2
\end{eqnarray}
In general, the root-mean-square contact distance $d$ is a function
of $R_g$ and $N$ ($d=d(R_g,N)$), and depends on compactness of a
chain.

\begin{figure}[h]
\includegraphics[width=9.0cm]{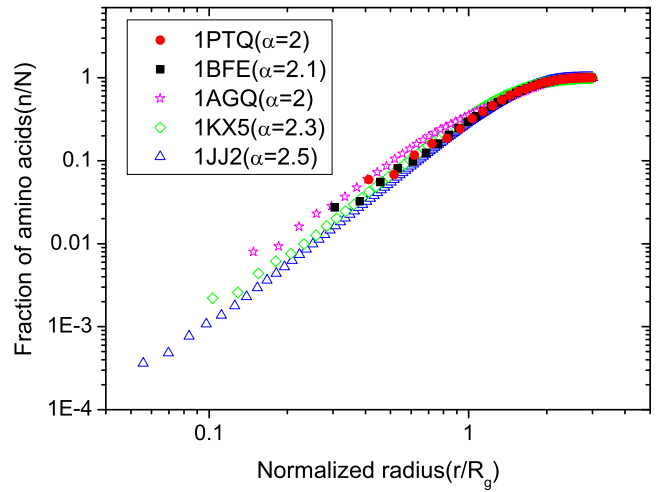}
\caption{A log-log plot for 5 illustrative proteins, which are
selected randomly from PDB except appropriate chain length.
($N=50,R_g=9.7$ for 1PTQ; $N=110,R_g=13.13$ for 1BFE;
$N=373,R_g=26.98$ for 1AGQ; $N=1268,R_g=38.74$ for 1KX5;
$N=6577,R_g=71.68$ for 1JJ2.) The data are obtained by counting the
number of amino acids ($n$) within different given radius ($r$)
starting at the geometrical center. It is clear that there exists a
wide self-similarity region between small-scale and large-scale
structure, with exponent $\alpha\approx2$. Beyond this region,
$\alpha$ drops to $1$ quickly, due to finite chain length.}
\end{figure}

As suggested by many authors, the protein can be regarded as a
fractal in some extent\cite{12,13,14,15,16,17}. If there exists a
self-similarity in number density between small-scale and
large-scale structure (Fig. 3), we can write
\begin{eqnarray}
(\bar{n}+1)/d^{\alpha}= N/R_g^\alpha
\end{eqnarray}
where $\alpha$ stands for fractal dimension of a protein's
structure. Thus the root-mean-square contact distance is obtained as
\begin{eqnarray}
d = (\bar{n}+1)^{1/\alpha} R_g/N^{1/\alpha}
\end{eqnarray}
Put into Eqn.(4),
\begin{eqnarray}
E_{\mathrm{ela}} = \frac{1}{4} \kappa \bar{n}(\bar{n}+1)^{2/\alpha}
\dfrac{R_g^2}{N^{2/\alpha - 1}}
\end{eqnarray}
Hence, the total energy is given by
\begin{equation}
 \label{eq:fe}
E= E_{\mathrm{rep}} + E_{\mathrm{ela}}  = k_B T v \frac{N^2}{R_g^3}
+ \frac{1}{4}\kappa \bar{n} (\bar{n} + 1)^{2/\alpha}
\dfrac{R_g^2}{N^{2/\alpha - 1}}
\end{equation}

In above argument, we neglect many other effects\cite{3,4,5,6}, such
as entropy effect ($E_{entropy}=\gamma R_g^2/N\propto N^{2\nu-1}$)
and three-body repulsive interaction ($E_{three}=\mu
N^3/R_g^6\propto N^{3-6\nu}$) etc. Nevertheless, in the region we
are interested ($\nu\in(\frac{1}{3},\frac{3}{5})$,
$\alpha\in(1,3)$), it is easily to check that
$\lim_{N\rightarrow\infty}E_{entropy}/E_{ela}=0,\;\lim_{N\rightarrow\infty}E_{three}/E_{rep}=0$.

In equilibrium state, radius of gyration can be estimated by
minimizing the total energy $E$. Let $\partial E/\partial R_g = 0$,
we have
\begin{equation}
 \label{eq:r}
R_g = \left[\frac{6 k_B T v}{\kappa \bar{n} (\bar{n} +
1)^{2/\alpha}}\right]^{1/5} N^{\frac{\alpha+2}{5\alpha}} \propto
N^{\frac{\alpha + 2}{5\alpha}}
\end{equation}
which gives
\begin{eqnarray}
\nu = \dfrac{\alpha+2}{5 \alpha}
\end{eqnarray}

In Fig.4, we can see that classical Flory's theory acts as an
extreme case in our new formula. In good solvent, polymer becomes
loose, and can be modeled as a one-dimensional long chain. Thus
$\alpha = 1$, which gives $\nu = 3/5$. In poor solvent, polymer is
highly compressed by solvent pressure, and becomes as well-packed as
crystals. It means $\alpha = 3$, then $\nu = 1/3$.

In the case of protein under physiological condition, we have
$\alpha\approx2$ (Fig.3), so $\nu\approx2/5$. It suggests that many
amino acid residues($\propto N$) are distributed at the surface of a
protein; and the interior is not so compact as what having been
thought before. This result is also supported by other
researches\cite{7,12,13,14,15,16,17}.

\begin{figure}[h]
\includegraphics[width=9.0cm]{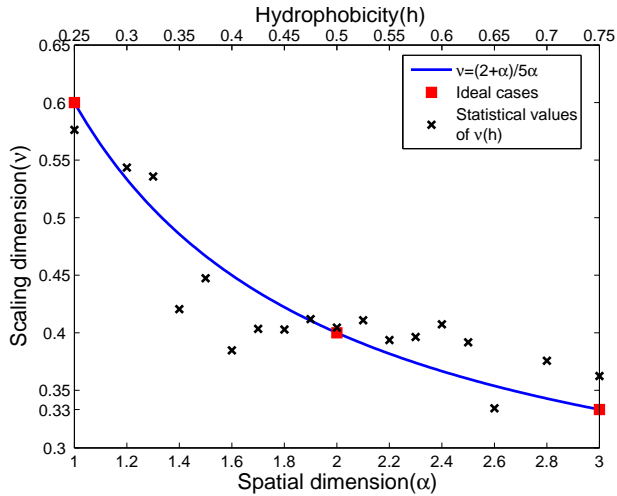}
\caption{\label{fig:epsart} The blue solid curve is for Eqn.(9). The
red squares are for three ideal cases respectively: polymer in good
solvent ($\alpha=1$), protein under physiological condition
($\alpha=2$) and polymer in poor solvent ($\alpha=3$). The crosses
stand for statistical values of exponent $\nu$ for proteins with
different hydrophobicity ($h$). $\nu(h)$ is estimated by
least-square linear fitting of uniformly selected statistical data
(proteins with same hydrophobicity within $\pm0.005$) from PDB. Data
for $h<0.25$ and $h>0.75$, as well as $h=0.275,0.675,0.725$ are
missing due to inadequate samples.}
\end{figure}

\section{Dependence on Hydrophobicity}
Above deduction is based on assumption of homopolymer. However, in
fact, protein is a heteropolymer made up of twenty different kinds
of amino acids. Thus exponent $\nu$ generally depends on the
component of the chain, especially its hydrophobicity. To study this
effect, a simple H-P model is introduced. Here we adopt the category
method of Kyte and Doolittle\cite{18}. All amino acids with positive
values in K-D method are regarded as hydrophobic (I, V, L, F, C, M,
A, G); while other ones with negative values are regarded as
hydrophilic (T, S, W, Y, P, H, E, N, Q, D, K, R).

The fraction of hydrophobic amino acids in a protein is defined as
its hydrophobicity($h$). Then if all amino acids are hydrophilic
($h=0$), which just corresponds to good solvent condition, the
protein will be fully extended with dimension $\alpha=1$. In this
case, constrains arising from covalent bonds are dominant
interaction against swelling tendency. If all amino acids are
hydrophobic ($h=1$), corresponding to poor solvent condition, the
protein is highly compressed by solvent pressure and dimension
$\alpha=3$. Strong hydrophobic interactions are balanced by excluded
volume effect between amino acid residues.

\begin{figure}[h]
\includegraphics[width=9.0cm]{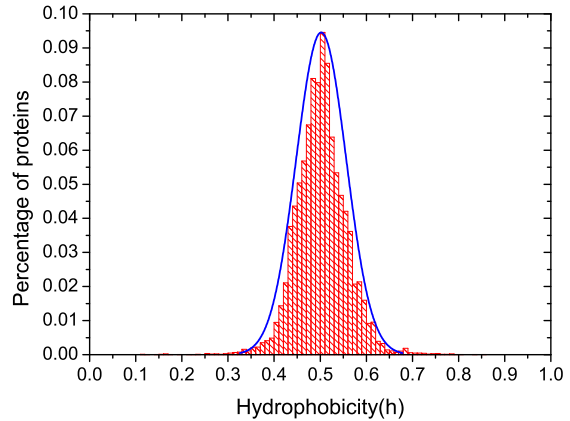}
\caption{\label{fig:epsart} Hydrophobicity distribution of 33901
protein structures in PDB. Data is fitted by a Gaussian curve
$Y=0.094\cdot e^{\frac{-(X-0.5)^2}{2*0.054^2}}$.}
\end{figure}

For natural proteins, their hydrophobicity has a Gaussian-like
distribution (Fig.5). In the region $h\in[0.4,0.6]$, scaling
exponent is almost unchanged (Fig.4), $\nu\approx2/5$. For $h<0.4$
or $h>0.6$, the number of natural proteins are quite limited. And
their corresponding exponent $\nu$ varies largely. Especially for
$h<0.25$ or $h>0.75$, the proteins can regarded as total hydrophilic
or hydrophobic respectively. These results hint appropriate
hydrophobicity is essential to maintain overall structure of a
natural protein.

To study how hydrophobicity affects scaling exponnet, we try to
theoretically predict $\alpha=\alpha(h),\; h\in[0,1]$.

\begin{figure}[h]
\includegraphics[width=0.5\textwidth,height=0.45\textwidth]{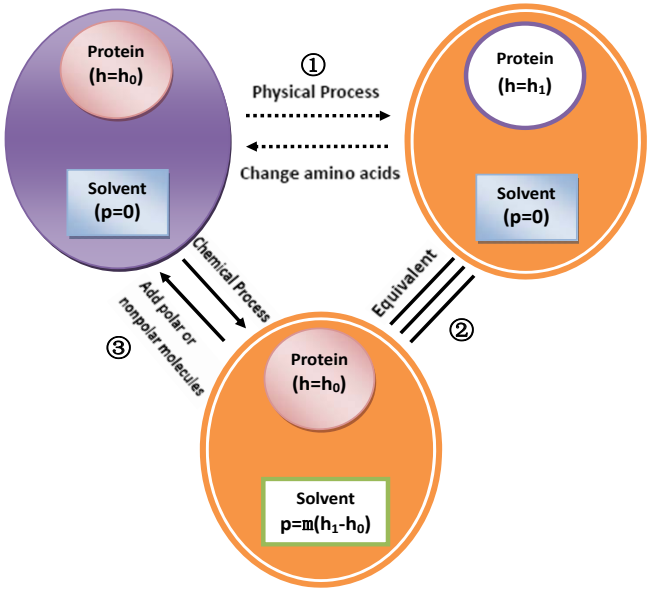}
\caption{\label{fig:epsart} Illustration for our main idea of
studying $\alpha(h)$. Process \textcircled{1}-\textcircled{3}
correspond to Eqn.(10)-(12) separately. Process \textcircled{1} is
what we want to study. However, varying the hydrophobicity of a
protein by amino acid mutation is not a chemical process, and hard
to grasp. Process \textcircled{2} is our main assumption: varying
the hydrophobicity of a protein is equivalent to varying the
polarity of solvent. Process \textcircled{3} is a real chemical
reaction. By adding polar or nonpolar molecules, we can control the
polarity of solvent.}
\end{figure}

We record the state of protein-solvent coupled system as
$X$\{Hydrophobicity of Protein, Polarity of Solvent\}$\equiv
X\{h,p\}$. Then two proteins with different hydrophobicity in same
water solution are written as
\begin{eqnarray}
X\{h=h_0,p=0\}\rightleftarrows X\{h=h_1,p=0\}
\end{eqnarray}
Here the polarity of water solution is set to zero. If we know how
above two states are changed into one another, we can predict the
relation of $\alpha(h)$. However, above process is connected by
amino acid mutation, which is not a chemical reaction and not easy
to analyze. Here we adopt an alternative way, which is based on the
assumption that varying the hydrophobicity of a protein is
equivalent to varying the polarity of solvent. According to
biochemistry, the hydrophobicity of a protein is closely related to
the polarity of solvent. The more polar the solvent is, the more
hydrophobic the protein is; while the less polar the solvent is, the
more hydrophilic the protein will be. Thus we can assume following
two systems are equivalent
\begin{eqnarray}
X\{h=h_1,p=0\}\equiv X\{h=h_0,p=m(h_1-h_0)\}
\end{eqnarray}
Here, we adopt a linear relationship between hydrophobicity and
polarity, and its validity remains to be verified by experiments.
From Eqn.(10) and (11), we can turn a physical process-varying
hydrophobicity of a chain by amino acid mutation, to a chemical
process-varying polarity of solvent by adding polar or nonpolar
molecules (Fig.6). Thus, Eqn.(10) is equivalent to following process
\begin{eqnarray}
X\{h=h_0,p=0\}\rightleftharpoons X\{h=h_0,p=m(h_1-h_0)\}
\end{eqnarray}

Now the study on $\alpha(h)$ is changed into a chemical reaction.
Suppose there are three separated stable thermal states
$X_1,X_2,X_3$, which represent proteins in good solvent, under
physiological condition and in poor solvent respectively. Their
corresponding fractal dimensions are $\alpha(X_1)=1$,
$\alpha(X_2)=2$ and $\alpha(X_3)=3$.

We start from the state under physiological condition. When the
condition is changed from water solution to good solvent, which can
be done by adding nonpoler molecules ($N$), proteins will change
from $X_2$ state to $X_1$ state, according to following chemical
process
\begin{equation}
X_2\rightleftharpoons^{k_1}_{k_{-1}} X_1
\end{equation}
Here reaction constants $k_1$ and $k_{-1}$ depend on the
concentration of nonpolar molecules added ($[N]$ is normalized to be
in $[0,1]$). Let $[X_i]$ be the fraction of proteins in state $X_i$.
When system reach equilibrium state, we have
\begin{eqnarray}
[X_1]/[X_2]=k_1/k_{-1}=K_1([N])
\end{eqnarray}
Here, a power function is chosen for above relation
\begin{eqnarray}
K_1([N])=C_1[N]^{m_1}
\end{eqnarray}
Due to the conservation law of matter $[X_1]+[X_2]=1$, we have
$[X_1]=K_1/(1+K_1),\;[X_2]=1/(1+K_1)$. Then for proteins with
solvent condition between water solution and good solvent, their
average fractal dimension is given by a Hill function
\begin{eqnarray}
\alpha=\alpha(X_1)[X_1]+\alpha(X_2)[X_2]=1+\frac{1}{1+C_1[N]^{m_1}}
\end{eqnarray}
with $[N]\in[0,1]$ and $C_1\gg1$.

Similarly, we can study proteins changed from $X_2$ state to $X_3$
state, or from water solution to poor solvent, which can be done by
adding poler molecules ($P$). This process is described as
\begin{equation}
X_2\rightleftharpoons^{k_2}_{k_{-2}} X_3
\end{equation}
Thus in the equilibrium state,
\begin{eqnarray}
[X_3]/[X_2]=k_2/k_{-2}=K_2([P])=C_2[P]^{m_2}
\end{eqnarray}
$[P]$ is the concentration of polar molecules added, and normalized
to be in $[0,1]$. As $[X_2]+[X_3]=1$,
$[X_2]=1/(1+K_2),\;[X_3]=K_2/(1+K_2)$. For proteins with solvent
condition from water solution to poor solvent, the average fractal
dimension is given by
\begin{eqnarray}
\alpha=\alpha(X_2)[X_2]+\alpha(X_3)[X_3]=3-\frac{1}{1+C_2[P]^{m_2}}
\end{eqnarray}
with $[P]\in[0,1]$ and $C_2\gg1$.

Take a linear relationship between hydrophobicity and polarity:
$h=\frac{1-[N]}{2}$ and $h=\frac{1+[P]}{2}$, we can fit the
statistical data by Eqn.(16) and (19) with appropriate values of
$m_1,\;m_2,\;C_1,\;C_2$ (Fig.7).

\begin{figure}[h]
\includegraphics[width=9.0cm]{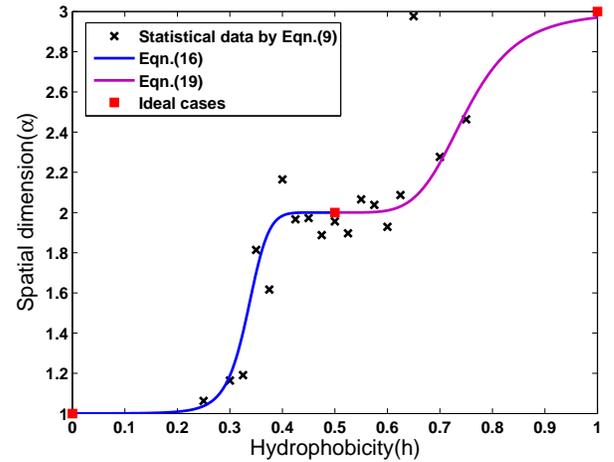}
\caption{\label{fig:epsart} Statistical data for $\alpha=\alpha(h)$,
which is obtained through inverse transform of data shown in Fig.4
by Eqn.(9). Fitting curve are given by Eqn.(16) and (19), with
$m_1=8,\;m_2=5,\;C_1=3^8,\;C_2=2^5$.}
\end{figure}

A suggested function of $\alpha=\alpha(h)$ is given by
\begin{eqnarray}
&&\alpha(h)= \left\{ \begin{array}{ll}
1+\frac{1}{1+3^8(1-2h)^{8}}, & \textrm{for h$\in$[0,0.5]}\\
3-\frac{1}{1+2^5(2h-1)^5}, & \textrm{for h$\in$(0.5,1]}\\
\end{array} \right.
\end{eqnarray}

\section{Conclusion}
In summary, we have derived a unified formula for the scaling law
between radius of gyration and the length of homopolymer chain. It
shows that this exponent is generally correlated with the fractal
dimension of a chain under certain solvent condition. Our new
formula covers the well-known Flory's theory for polymers under good
and poor solvent conditions as two extreme cases. It can be applied
to proteins under physiological condition ($\nu\approx2/5$) too,
with a predicted fractal dimensional $\alpha\approx2$. Influence of
hydrophobicity on the compactness of a protein has also been studied
through a simple H-P model. By considering the equivalence between
protein-solvent coupled systems, we turn a physical process-varying
the hydrophobicity of a chain by amino acid mutation, to a chemical
process-varying the polarity of solvent by adding polar or nonpolar
molecules. This enables us to derive a functional relation between
hydrophobicity and fractal dimension, with reasonable agreement to
statistical data. This relation will be helpful for protein
structure prediction. Our results indicate that the protein may
share the same basic principle with homopolymer, despite its
speciality as a heteropolymer. Hope this work can shed light on the
mechanism of protein folding and stability of protein structures.

\begin{acknowledgments}
The authors thank Professor C.C.lin  and Professor Kerson Huang for
their guidance and many useful comments. Thank Professor Wen-An Yong
and Doctor Weitao Sun for their helpful discussions.
\end{acknowledgments}

\appendix

\section{Direct study of $\nu(h)$}
Although we can get $\nu(h)$ according to Eqn.(9) and (20), a direct
prediction is also possible in the same way as Sec.III. Suppose
$\nu(X_1)=3/5,\nu(X_2)=2/5,\nu(X_3)=1/3$, the average scaling
exponent is given by
\begin{eqnarray}
&&\nu=\nu(X_1)[X_1]+\nu(X_2)[X_2]\nonumber\\
&&\quad=\frac{3}{5}-\frac{1}{5(1+C_1[N]^{m_1})}\nonumber\\
&&\quad=\frac{3}{5}-\frac{1}{5[1+C_1(1-2h)^{m_1}]}
\end{eqnarray}
for $h\in[0,0.5]$, $C_1\gg1$; and
\begin{eqnarray}
&&\nu=\nu(X_2)[X_2]+\nu(X_3)[X_3]\nonumber\\
&&\quad=\frac{1}{3}+\frac{1}{15(1+C_2[P]^{m_2})}\nonumber\\
&&\quad=\frac{1}{3}+\frac{1}{15[1+C_2(2h-1)^{m_2}]}
\end{eqnarray}
for $h\in(0.5,1]$, $C_2\gg1$.

\begin{figure}[h]
\includegraphics[width=9.0cm]{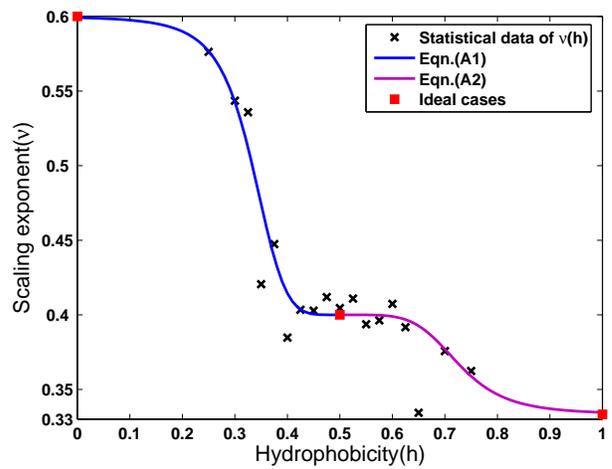}
\caption{\label{fig:epsart} The statistical data is the same as
Fig.4. Fitting curves are given by Eqn.(A1) and (A2), with
$m_1=m_2=5,\;C_1=3^5,\;C_2=2.2^5$.}
\end{figure}

A suggested function of $\nu=\nu(h)$ is given by
\begin{eqnarray}
&&\nu(h)= \left\{ \begin{array}{ll}
\frac{3}{5}-\frac{1}{5[1+3^5(2h-1)^5]}, & \textrm{for h$\in$[0,0.5]}\\
\frac{1}{3}+\frac{1}{15[1+2.2^5(2h-1)^5]}, & \textrm{for h$\in$(0.5,1]}\\
\end{array} \right.
\end{eqnarray}

\end{document}